\documentclass[11pt]{article}
\pdfoutput=1

\usepackage{cite}
\usepackage{booktabs}
\usepackage{amsmath,amssymb,amsbsy,amstext, amsthm, simplewick}
\usepackage{hyperref}
\usepackage{graphicx}
\usepackage{amsfonts}
\usepackage{amssymb}
\usepackage[small]{caption}
\usepackage{upgreek}
\usepackage[svgnames,dvipsnames,x11names,table]{xcolor}
\usepackage{multirow} 
\usepackage{geometry}
\usepackage[hang,flushmargin]{footmisc}
\usepackage{bm}
\usepackage{braket}
\usepackage{subcaption}
\usepackage{simpler-wick}
\usepackage{mathtools}

\usepackage{colortbl}

\makeatletter
\newlength{\apb@width}
\newcommand{\autoparbox}[2][c]{\settowidth{\apb@width}{#2}\parbox[#1]{\apb@width}{#2}}

\makeatother

\definecolor{lightgray}{gray}{0.9}

\usepackage[framemethod=default]{mdframed}
\newmdenv[skipabove=7pt,
skipbelow=7pt,
rightline=false,
leftline=false,
topline=false,
bottomline=false,
backgroundcolor=gray!10,
linecolor=gray,
innerleftmargin=5pt,
innerrightmargin=5pt,
innertopmargin=5pt,
innerbottommargin=5pt,
leftmargin=0cm,
rightmargin=0cm,
linewidth=4pt]{eBox}

\numberwithin{equation}{section}

\def\beq{\begin{equation}}
\def\eeq{\end{equation}}

\def\bea{\begin{eqnarray}}
\def\eea{\end{eqnarray}}

\def\beq{\begin{equation}}
\def\eeq{\end{equation}}
\def\bea{\begin{eqnarray}}
\def\eea{\end{eqnarray}}

\def\H{{\cal H}}

\def\hs{{\hat {\boldsymbol{s}}}}

\def\k{{\vec k}}

\def\q{{\vec q}}

\def\p{{\vec p}}

\def\x{{\vec x}}

\def\y{{\vec y}}

\def\z{{\vec z}}

\DeclareRobustCommand{\SkipTocEntry}[4]{}

\def\zk{| 0 \rangle_{\boldsymbol{s}}}
\def\ok{| 1 \rangle_{\boldsymbol{s}}}

\definecolor{blue3}{RGB}{31, 119, 180}
\definecolor{red3}{RGB}{	214, 39, 40}
\definecolor{orange3}{RGB}{255, 127, 14}
\definecolor{green3}{RGB}{44, 160, 44}

\begin{document}

\begin{titlepage}
\setcounter{page}{1} \baselineskip=15.5pt 
\thispagestyle{empty}

\begin{center}
{\fontsize{18}{18} \bf A Flat Space Analogue for \\[5pt] the Quantum Origin of Structure }
\end{center}

\vskip 20pt
\begin{center}
\noindent
{\fontsize{12}{18}\selectfont  Daniel Green and Yiwen Huang }
\end{center}

\begin{center}
\vskip 4pt
\textit{ Department of Physics,University of California at San Diego, \\ La Jolla, CA 92093, USA}

\end{center}

\vspace{0.4cm}
 \begin{center}{\bf Abstract}
 \end{center}
 
 \noindent  The analytic structure of non-Gaussian correlators in inflationary cosmologies has recently been proposed as a test of the quantum origin of structure in the universe.  To further understand this proposal, we explore the analogous equal-time in-in correlators in flat space and show they exhibit the same features as their cosmological counterparts. The quantum vacuum is uniquely identified by in-in correlators with a total energy pole and no additional poles at physical momenta.  We tie this behavior directly to the S-matrix and show that poles at physical momenta always arise from scattering of particles present in the initial state.  We relate these flat-space in-in correlators to the probability amplitude for exciting multiple Unruh-de Witt detectors.  Localizing the detectors in spacetime, through the uncertainty principle, provides the energy and momentum needed to excite the vacuum and explains the connection to cosmological particle production.  In addition, the entanglement of these detectors provides a probe of the entangled state of the underlying field and connects the properties of the correlators to the range of entanglement of the detectors.  
 
\end{titlepage}

\restoregeometry

\newpage
\setcounter{tocdepth}{2}
\tableofcontents

\newpage

\section{Introduction}

Quantum mechanics is responsible for a number of physical phenomena that are impossible in classical physics.  Characterizing these unique properties of quantum systems is a problem of wide scientific interest. Bell famously provided one such example~\cite{Bell:1964kc} where the measurements of a small number of entangled spins (qubits) cleanly distinguishes quantum from local classical physics~\cite{Bohm:1951xw,Bohm:1951xx}.  Physical properties of real world systems are far more complex and therefore isolating their uniquely quantum behavior can be more challenging~\cite{Vidal:2002zz,Lu:2018yxh,Lu:2019xwg}.  One particularly interesting example is the origin of the initial density fluctuations in the universe. They are believed to have arisen from quantum fluctuations during an inflationary epoch~\cite{Mukhanov:1981xt,Hawking:1982cz,Guth:1982ec,Starobinsky:1982ee,Bardeen:1983qw}, however, this hypothesis has been difficult to test observationally. The universe is classical on cosmological scales and one cannot easily apply Bell's inequality to the density fluctuations directly~\cite{Starobinsky:1986fx,Grishchuk:1990bj} (but see e.g.~\cite{Campo:2005sv,Lim:2014uea,Maldacena:2015bha,Martin:2015qta,Goldstein:2015mha,Nelson:2016kjm,Choudhury:2016cso,Martin:2017zxs,Shandera:2017qkg,dePutter:2019xxv,Martin:2019oqq,Brahma:2021mng,Gomez:2021yhd,Martin:2021qkg,Espinosa-Portales:2022yok} for ongoing work).  Instead, one is lead to ask if quantum mechanics {\it was} important in establishing the (statistical) initial conditions for our classical cosmological observations.

\begin{figure}[h!]
\centering
\includegraphics[width=5in]{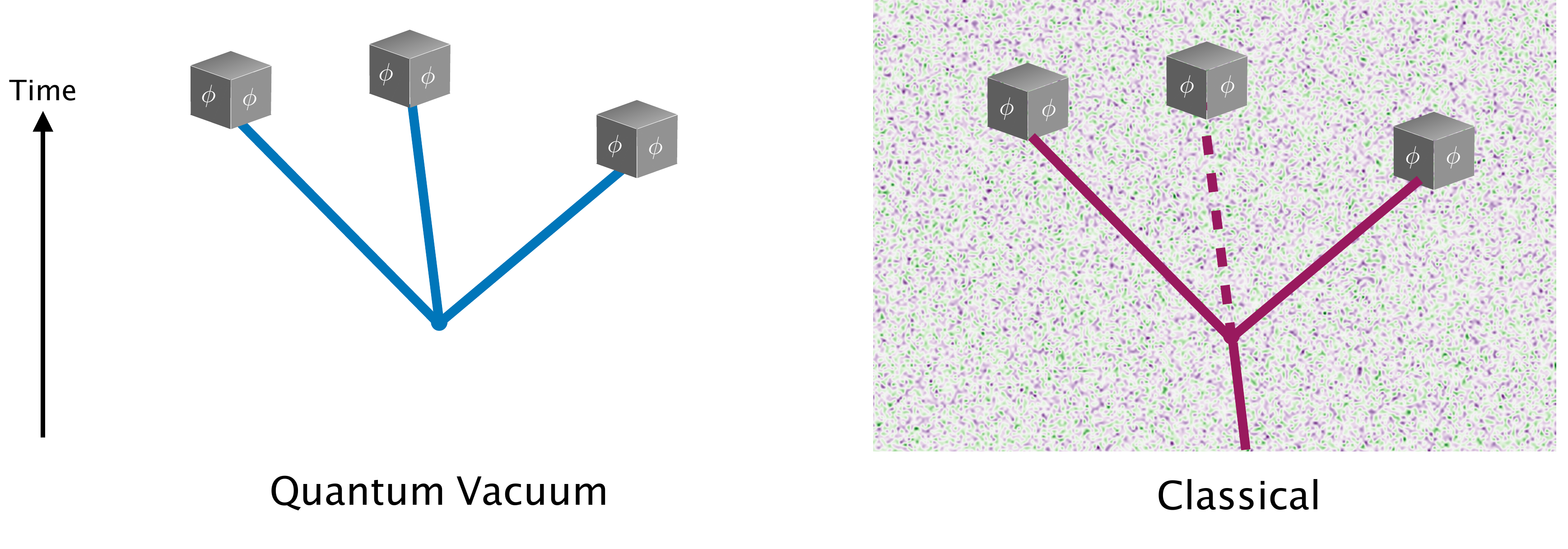}
\caption{Illustration of the difference between non-Gaussian in-in correlations of $\phi$ for quantum vacuum fluctuations (left) and classical fluctuations (right).  Measuring quantum fluctuations of $\phi$ at three spacelike separated points corresponds to the creation of three particles from the vacuum, producing a total energy pole in the correlator.  In contrast, classical fluctuations only occur in a state containing particles.  Any local classical process that produces a total energy pole will also cause particles in the initial state to decay, producing additional three point correlations with poles at physical momenta.}
\label{fig:Q_vs_C}
\end{figure}

Quantum effects play a crucial role in the dynamics of the early universe in many inflationary models.  While the resulting observational signals can often be traced to quantum mechanics, the challenge is showing no classical mechanism could produce the same signal~\cite{Berera:1995ie,Berera:1998px,Green:2009ds,LopezNacir:2011kk,LopezNacir:2012rm,Turiaci:2013dka}.  One proposal of this kind was made in~\cite{Green:2020whw}, where it was shown that the analytic structure of (non-Gaussian) correlation functions is different for classical and quantum theories when the correlations are produced by local evolution. The origin of this difference arises from the non-zero number of particles needed to produce classical density fluctuations, as illustrated in Figure~\ref{fig:Q_vs_C}.  Creation of particles from the quantum vacuum violates energy conservation but is allowed because of the uncertainty principle. In contrast, physical particles will scatter and decay in an interacting theory while conserving energy, giving rise to poles at physical momentum for classical fluctuations.  These results are consistent with a number of results relating scattering to the analytic structure of cosmological correlators~\cite{Maldacena:2011nz,Raju:2012zr,Arkani-Hamed:2015bza,Lee:2016vti,Arkani-Hamed:2018kmz,Arkani-Hamed:2018bjr,Benincasa:2018ssx,Baumann:2019oyu,Pajer:2020wxk,Baumann:2020dch,Bonifacio:2021azc,Cabass:2021fnw,Baumann:2021fxj,Baumann:2022jpr}.  These features of the correlators are also directly tied to the prospects of observing the signal~\cite{Gleyzes:2016tdh,Flauger:2013hra,Baumann:2021ykm}.  

The relationship between correlators and scattering is, of course, best understood in flat space. The LSZ reduction formula~\cite{Lehmann:1954rq} gives a rigorous map between in-out correlators and S-matrix elements.  While measurements of flat space correlators are not subject to the same limitations as cosmology, we can still ask if the analytic structure of flat space in-in correlators encodes the quantum vacuum in the same way.  Furthermore, one could hope to use LSZ to connect the difference between the analytic structure of classical and quantum correlators directly to scattering of particles in the initial state.

On a purely theoretical level, flat space provides a testing ground for our understanding of cosmological correlators.  Yet, if in-in correlators are encoding the physics of the quantum vacuum, one would naturally like to understand how they are related to measurable quantities.  Cosmological expansion is essential in producing fluctuations from the vacuum during inflation and does not occur in flat space.  To make sense of their flat space analogues, we must introduce particle detectors localized in the spacetime (Unruh-de Witt detectors~\cite{Unruh:1976db,DeWitt:1980hx}).  The very act of measuring the state of the quantum field at a localized point in spacetime introduces the energy and momentum needed to excite the vacuum (breaking the time and space translations). This observation was essential for making sense of Unruh radiation~\cite{Unruh:1976db} (i.e.~the Rindler temperature), another example of particle production in flat space.  In that case, a thermal distribution of particles is seen by a constantly accelerating (Rindler) observer because of the energy and momentum needed to accelerate the detector in the first place~\cite{Unruh:1983ms,Kaplanek:2019dqu}.   Similarly, measuring correlators at a fixed time also requires the injection of energy (by the uncertainty principle) and naturally explains the apparent particle production from nothing which is encoded in the in-in correlators.

The non-local correlations of a field in the quantum vacuum also generates entanglement between the various particle detectors used to detect them~\cite{VALENTINI1991321,Reznik:2002fz,Reznik:2003mnx,Lin:2007mu,Martin-Martinez:2012chf,Nambu:2013rta,Salton:2014jaa,Hummer:2015xaa,Pozas-Kerstjens:2015gta,Sachs:2017exo,Henderson:2018lcy}.  As such, the detector entanglement represents a probe of the entanglement of the interacting vacuum of the fields themselves.  Famously, the entanglement entropy of the fields on a finite region is expected to be proportional to the area for the quantum vacuum~\cite{Srednicki:1993im,Eisert:2008ur} and the volume for a generic excited state (see e.g.~\cite{Casini:2022rlv} for review).  While the entanglement entropy is not a quantity we can easily measure (or calculate), naturally one would like to understand if the non-Gaussian signature of the quantum vacuum state is related. 

In this paper, we will expose the connection between flat space scattering, entanglement, and cosmological observables through the properties of in-in correlators.  While they are a less natural observables in flat space than the S-matrix, we show that there is a precise link between the structure of poles in the in-in correlators and the associated scattering processes, as illustrated in Figure~\ref{fig:Q_vs_C}.  This provides a robust demonstration that the poles appearing at physical momenta in classical states are directly tied to the decay or scattering of particles in the initial state.  The poles are absent in the quantum vacuum because it contains no particles, connecting the analysis of~\cite{Green:2020whw} to flat space amplitudes.  

To make physical sense of these results, we show that the in-in correlators can be interpreted as an the amplitudes to excite multiple UdW detectors localized at space-like separated points. These detectors then provide a natural connection between cosmological Bell-type tests and more typical characterizations in terms of entanglement.  The entanglement of these detectors shares many similarities with the entanglement of the underlying field; yet, we can directly connect these properties to the underlying in-in correlators.   We will see that the analytic structure of the correlators is directly related to short or long ranged entanglement of the detectors.

This paper is organized as follows: in Section~\ref{sec:Smatrix}, we will discuss the relationship between the S-matrix and the in-in correlators, demonstrating our main results about the analytic structure of these correlators.  In Section~\ref{sec:detector}, we show how to interpret our results in terms of UdW detectors.  We then show how these detectors are entangled in Section~\ref{sec:entanglement}, and conclude in Section~\ref{sec:conclusions}.  Appendix~\ref{app:nonlocal} contains additional details about the relationship between enforcing causality of our classical theory and the existence of anti-particles.

\section{From In-In Correlators and the S-matrix} \label{sec:Smatrix}

The process of creating particles from the vacuum is an inherently quantum mechanical phenomenon.  It gives rise to structure in inflationary models~\cite{Mukhanov:1981xt,Hawking:1982cz,Guth:1982ec,Starobinsky:1982ee,Bardeen:1983qw} and Hawking radiation from black holes~\cite{Hawking:1975vcx}. In flat space, such a process is forbidden by energy conservation, but the amplitude is still formally well-defined as it is related to physical scattering processes by crossing symmetry.  This statement can be made rigorously through the LSZ reduction formula~\cite{Lehmann:1954rq}.  

We would like to understand how quantum fluctuation can be distinguished from classical (e.g. thermal) fluctuations.  Classical fluctuations may occur in any spacetime and thus we can ask this question in flat space as well. We will show in this section that an isolated total energy pole in an equal-time correlator (in-in or in-out) is precisely a reflection of the amplitude for production of particles from the vacuum.  We will then show that for classical fluctuations, additional poles arise from the on-shell scattering processes of particles in the initial state.  This difference between quantum vacuum fluctuations and classical fluctuations is illustrated in Figure~\ref{fig:Q_vs_C}. 

\subsection{The In-In Formalism} \label{sec:inin}
Cosmological correlators are described by (equal time) in-in correlations functions.  In perturbation theory, these are defined as~\cite{Weinberg:2005vy,Weinberg:2006ac}
\beq\label{eq:inin}
\langle {\rm in}|  Q(t) |{\rm in} \rangle=\left\langle\bar{T} \exp \left[i \int_{-\infty(1+i \epsilon)}^{t} H_{\mathrm{int}}(t') d t' \right]  \, Q_{\mathrm{int}}(t)  \, T \exp \left[-i \int_{-\infty(1-i \epsilon)}^{t} H_{\mathrm{int}}(t') d t' \right]\right\rangle \ ,
\eeq
where $H_{\mathrm{int}}(t) = \int d^3 x \sqrt{-g} \H_{\rm int} (\x,t)$, $g$ is the determinant of the metric, $\H_{\rm  int}(\x, t)$ is the Hamiltonian density of the interaction Hamiltonian, and $Q_{\rm int}(t)$ is the operator $Q(t)$ in terms of the interaction picture fields.  Computed in a quasi-de Sitter background for super-horizon modes (i.e.~points separated by super-horizon distances or fourier modes with super-horizon wavelengths) the in-in correlators give the classical statistical correlations of the initial density fluctuations.

It is occasionally useful to express the in-in correlators in terms of commutators (although, technically, it only  applies when $\epsilon  = 0$).  Expanding the time-ordered exponentials, one finds
\begin{subequations}
\begin{align}
\langle {\rm in}|  Q(t) |{\rm in} \rangle &= \sum_{N= 0}^{\infty} i^{N} \int_{-\infty}^{t} d t_{N} \int_{-\infty}^{t_{N}} d t_{N-1} \cdots \int_{-\infty}^{t_{2}} d t_{1} \nonumber \\[4pt] 
&\hspace{60pt} \times\Big\langle\big[H_{{\rm int}}\big(t_{1}\big) \big[H_{{\rm int}}\big(t_{2}\big)  \cdots\big[H_{{\rm int}}\big(t_{N}\big), Q_{\mathrm{int}}(t) \big] \cdots\big]\big]\Big\rangle  \ . \label{eq:commutator}
\end{align}
\end{subequations}
This representation is useful for two reasons.  First, this expression makes causality manifest as the commutators must vanish outside the lightcone.  Second, the non-zero commutator is the defining characteristic of a quantum theory and thus this representation is useful in isolating the quantum nature of the correlators.  

Our goal in this paper is to understand what aspects of the in-in correlator reflect truly quantum fluctuations and what other aspects could arise purely classically.  With this in mind, we will focus on the fluctuations of a single scalar field $\phi$, which may be represented as a quantum mechanical operator or a classical stochastic variable.  Following~\cite{Green:2020whw}, we can describe the free classical or quantum theories using the mode expansion
\beq
\phi(\x,t) = \int \frac{d^3 k}{(2\pi)^3} e^{i \k\cdot \x}  \frac{1}{\sqrt{2k}} [ a^\dagger_{-\k}   \, e^{i k t} +a_{\k} e^{-i k  t} ] \ ,
\eeq
where $k\equiv |\vec k|$.  The distinction between quantum and classical is how we interpret $a_\k$ and $a^{\dagger}_\k$.

In the quantum theory, $a_\k$ and $a_\k^{\dagger}$ are operators satisfying 
 \beq
 [a_\k, a^\dagger_{\k'}] = (2\pi)^3 \delta(\k - \k')  \qquad a_\k |0 \rangle = 0\,,
  \eeq
where $|0 \rangle$ is the vacuum of the free theory.  In contrast, in the classical theory they are only random variables obeying the statistics
\beq
\langle a^{\dagger}_\k a_{\k'}\rangle_c =\frac{1}{2} (2\pi)^3  \delta(\k-\k') = \langle  a_{\k'} a^{\dagger}_\k \rangle_c\, \ .
 \eeq
In the free theory, these choices give the same equal-time correlators which are (essentially) the only cosmological observable.  Of course, in flat space, we would be free to directly measure the commutators of the operators or non-equal time correlators to expose the difference between classical and quantum mechanics, but we will restrict ourselves to equal-time to parallel the cosmological correlators.  

We can start with a simple example for illustration: given a massless scalar $\phi$ with a cubic self interaction, $\H_{\rm int} = \frac{1}{3!} \mu \phi^3$, the in-in three-point function in terms of fourier modes (the bispectrum) is given by 
\bea
\langle \phi(t, \k_1) \phi(t,\k_2) \phi(t,\k_3) \rangle &=& 2 {\rm Im} \int^t_{-\infty} dt'\frac{\mu}{8 k_1 k_2 k_3} e^{- i (k_1+k_2+k_3) (t-t')}\\
&=& - \frac{\mu}{4 k_1 k_2 k_3 (k_1+k_2+k_3) } \label{eq:vac_bi} , 
\eea
where $\langle Q(t) \rangle \equiv \langle \Omega | Q(t) | \Omega \rangle$ is the correlation in the interacting vacuum, $|\Omega \rangle$.  We have only fourier transformed the spatial coordinates by analogy with a typical cosmological correlator. Just like a cosmological correlator, we see that this in-in correlation functions exhibits a pole only in the total energy $k_t= (k_1+k_2+k_3)$.  The presence of such a pole is a unique signature of the quantum vacuum and therefore we would like to better understand the physical significance of this correlation.

\vskip 5pt
For classical fluctuations, the appearance of additional poles can be seen by perturbatively solving the equations of motion with the same cubic interaction, $\H_{\rm int} = \frac{1}{3!} \mu \phi^3$, such that
\beq
\phi^{(2)}(\k,t) =\frac{\mu}{2} \int^t dt'  G(k; t-t') \int  \frac{d^3 p}{(2\pi)^3} \phi(\p, t') \phi(\k-\p, t')
\eeq
where $G(k,t) =\sin k t /k$ is the causal Green's function. Using this to calculate the bispectrum we find 
\beq
\langle \phi(t, \k_1) \phi(t,\k_2) \phi(t,\k_3) \rangle_c =  \frac{\mu}{16 k_1 k_2 k_3} \left(\frac{3}{ k_t } + \sum_{i=1}^3\frac{1}{k_t-2 k_i} \right) \ ,
\eeq
where we assumed the contribution from $t \to -\infty$ vanishes\footnote{For vacuum correlators, this is equivalent to the $i\epsilon$ prescription.  For excited states, this will ultimately be tied to how the poles at physical momentum are resolved.}.  We see that the classical example has poles both in the total energy and in folded configurations where $k_1 = k_2+k_3$ and permutations therefore.  In this respect, we see that flat space in-in correlators exhibit the same structure as the cosmological counter-parts~\footnote{In~\cite{Green:2020whw}, it was shown that this conclusion is an inevitable consequence of causality and Lorentz invariance.  In Appendix~\ref{app:nonlocal}, we extend this argument to show that it is equivalent to the need for anti-particles in a relativistic quantum theory.}.  

Naturally, we would also like to understand how one interpolates between the quantum mechanical and classical result.  After all, we certainly live in a quantum universe and would like to understand how classical fluctuations would arise.  We can make this connection by taking the quantum theory in the limit where every momentum state is highly occupied,
\beq \label{eq:n_state}
|n_{\k} \rangle =\frac{1}{\sqrt{n!} } \left( a^{\dagger}_{\k}\right)^n |0\rangle \quad  \to \quad |n \rangle \equiv \bigotimes_{\k_i} |n_{\k} \rangle . 
\eeq
Suppose we have a real field $\phi$ is in an $n$-particle state, $|n \rangle$, we see that the operator $\hat \phi$ acts schematically
\bea
 \phi(\k,t) |n \rangle   &\to& \sqrt{n_{-\k}+1} f(k,t) |n_{-\k}+1\rangle |\hat n;  -\k \rangle + \sqrt{n_{\k}} f^*(k,t) |n_{\k}-1\rangle |\hat n;  \k \rangle\,\\
  \langle n |  \phi(\k,t)   &\to&  \sqrt{n_{\k}+1} f^*(k,t) \langle n_{\k}+1 |  \langle \hat n;  \k | + \sqrt{n_{-\k}} f(k,t) \langle n_{-\k}-1 |\langle \hat n; - \k | ,
\eea
where $f(k,t) = e^{i k t} /\sqrt{2k}$ is the positive frequency classical solution, and we defined
\beq
|\hat n; \q \rangle \equiv |n \rangle \equiv \bigotimes_{\k_i \neq \q} |n_{\k_i} \rangle \ .
\eeq
The two point function in this state is therefore
\beq
\langle \phi(\k, t)  \phi(\k',t') \rangle =   \left( (n+1)f^*(k,t) f(k',t') +  n f(k,t) f^*(k',t') \right) (2\pi)^3 \delta(\k+\k') \ ,
\eeq
which reproduces our classical statistical when taking $n \to \infty$ with $f(\k,t) \sqrt{n}$ fixed.  In this limit, the two point function becomes
\beq
\langle \phi(\k, t)  \phi(\k',t') \rangle \to n \left(f(k,t) f^*(k',t')+f^*(k,t) f(k',t') \right) (2\pi)^3 \delta(\k+\k') \ .
\eeq
This expression is symmetric in $t \leftrightarrow t'$ and thus shows that $\phi(\k,t)$ and $\phi(\k',t')$ commute, as we would expect for a classical variable and not for a quantum mechanical operator.

\subsection{Relation to the S-Matrix}

Now we want to understand how the poles in our in-in correlators are related to physical scattering processes.

\subsubsection*{Quantum Vacuum}

The most direct relationship between scattering and correlation functions is the LSZ reduction formula~\cite{Lehmann:1954rq}. Given a time-ordered (quantum) vacuum correlation function in flat space, we can extract the associated S-matrix elements via
\begin{align}
\left\langle \{ p_{j} \}_{\text{out}}\right| \{ q_{i} \}_{\text{in}}\rangle=&\int \prod_{i=1}^{m}\left\{\mathrm{d}^{4} x_{i} \frac{i e^{i q_{i} \cdot x_{i}}\left(-\square_{x_{i}}+m^{2}\right)}{(2 \pi)^{\frac{3}{2}} Z^{\frac{1}{2}}}\right\} \prod_{j=1}^{n}\left\{\mathrm{d}^{4} y_{j} \frac{i e^{-i p_{j} \cdot y_{j}}\left(-\square_{y_{j}}+m^{2}\right)}{(2 \pi)^{\frac{3}{2}} Z^{\frac{1}{2}}}\right\} \nonumber \\
& \times \left\langle \Omega\left|\mathrm{T} \phi\left(x_{1}\right) \ldots \phi\left(x_{m}\right) \phi\left(y_{1}\right) \ldots \phi\left(y_{n}\right)\right| \Omega\right\rangle \ ,
\end{align}
where we are using the metric signature $(-+++)$. The scattering states that appear on the left of this expression are defined by
\beq
| \{ q_{i} \}_{\text{in}}\rangle = \lim_{t\to -\infty} \prod_i \sqrt{2\omega_{\q_i}} a^\dagger_{\q_i} |\Omega\rangle \qquad \left\langle \{ p_{j} \}_{\text{out}}\right| = \lim_{t\to +\infty}  \langle \Omega | \prod_j \sqrt{2\omega_{\p_j}} a^\dagger_{\p_j} \ .
\eeq
Because the vacuum, $|\Omega\rangle$, is annihilate by $a_\k$, isolating the positive frequency via the Fourier transform (i.e.~integrating the correlation function with $\int dt e^{-i \omega t}$ for $\omega > 0$) also isolates a particle in the initial state.  

In perturbation theory, the time-ordered and in-in correlators are closely related, allowing us to directly related the poles in each to the associated S-matrix elements.  To calculate the in-out correlators, we need the time ordered Green's function
\beq
\langle 0| T(\phi(t_1,\k) \phi(t_2,\k') |0 \rangle = \frac{1}{2 k} e^{-i k |t_1-t_2|} \, (2\pi)^3 \delta(\k+\k') \ .
\eeq
We can calculate an equal time in-out correlator for our example with a cubic interaction to find the same result as the in-in correlator
\bea \
\langle T \phi(0, \k_1) \phi(0,\k_2) \phi(0,\k_3) \rangle &=& - i \int^{\infty}_{-\infty} dt \frac{\mu}{8 k_1 k_2 k_3} e^{- i (k_1+k_2+k_3)|t| }\\
&=& - \frac{\mu}{4 k_1 k_2 k_3 (k_1+k_2+k_3) } \ . 
\eea
To see the connection to the S-matrix elements, we need to consider unequal times such that
\begin{align}
\langle T \phi(t_1, \k_1) \phi(t_2,\k_2) \phi(t_3,\k_3) \rangle =& - i \int^{\infty}_{-\infty} dt \frac{\mu}{8 k_1 k_2 k_3} e^{-i k_1 |t_1-t|  } e^{-i k_2 |t_2-t|  }e^{-i k_3 |t_3-t|  }\\
=& - \frac{\mu}{8 k_1 k_2 k_3} \bigg(\frac{e^{-i k_2 (t_2-t_1) - i k_3(t_3-t_1) } }{k_1+k_2+k_3} \\
& + \frac{e^{i k_1 (t_1-t_2)  -i k_3 (t_3-t_2)} - e^{-i k_2 (t_2-t_1) - i k_3(t_3-t_1) } }{-k_1+k_2+k_3}  \\
& + \frac{e^{i k_1 (t_1-t_3)  + i k_2 (t_2-t_3)} - e^{i k_1 (t_1-t_2) - i k_3(t_3-t_2) } }{-k_1-k_2+k_3}  \\
& + \frac{e^{i k_1 (t_1-t_3)  + i k_2 (t_2-t_3)} }{k_1+k_2+k_3} \bigg) \ ,
\end{align}
where we have assume $t_1<t_2<t_3$ without loss of generality. 

Since our above expression assumes $t_1 < t_2 < t_3$, applying LSZ while maintaining this order is consistent if $t_1$ is associated with the initial state and $t_2, t_3$ are the final states.  We can calculate the S-matrix elements by first taking the Fourier transform,
\beq
\langle T \phi(\omega_3,\vec{k}_3)\phi(\omega_2,\vec{k}_2)\phi(\omega_1,\vec{k}_1)\rangle^{\prime} = \int dt_1 dt_2 dt_3 e^{i (-\omega_1 t_1+\omega_2 t_2+\omega_3 t_3)}\langle T \phi(t_3,\vec{k}_3)\phi(t_2,\vec{k}_2)\phi(t_1,\vec{k}_1)\rangle \ .
\eeq
Performing the integral imposing $t_1 < t_2 < t_3$ gives the term of interest
\beq\label{eq:inin_fourier}
\begin{aligned}
&\langle T \phi(\omega_3,\vec{k}_3)\phi(\omega_2,\vec{k}_2)\phi(\omega_1,\vec{k}_1)\rangle^{\prime} \\ &\supset \frac{i\mu}{8k_2k_2k_3}2\pi\delta(\omega_2+\omega_3-\omega_1)\Big(-\frac{1}{\omega_3-k_3}\frac{1}{\omega_3-k_3+\omega_2-k_2}\frac{1}{k_1+k_2+k_3}-\\&\frac{1}{\omega_3-k_3}\frac{1}{\omega_3+\omega_2-k_1}\frac{1}{-k_1+k_2+k_3}+\frac{1}{\omega_3-k_3}\frac{1}{\omega_3-k_3+\omega_2-k_2}\frac{1}{-k_1+k_2+k_3}-\\&
\frac{1}{\omega_3-k_1-k_2}\frac{1}{\omega_3-\omega_2-k_1}\frac{1}{-k_1-k_2+k_3}+\frac{1}{\omega_3-k_3}\frac{1}{\omega_2+\omega_3-k_1}\frac{1}{-k_1-k_2+k_3}\\&-\frac{1}{\omega_3-k_1-k_2}\frac{1}{\omega_3+\omega_2-k_1}\frac{1}{k_1+k_2+k_3}\Big) \ .
\end{aligned}
\eeq
Now we want to isolate the part of the correlator that encodes the $\k_1 \to \k_2+\k_3$ scattering amplitude.  Using LSZ, we see 
\beq
\begin{aligned}
\langle k_2,k_3|k_1\rangle &= \lim_{\omega_i \to k_i}(\omega_1-k_1)(\omega_2-k_2)(\omega_3-k_3)(8 k_1 k_2 k_3) \langle T \phi(\omega_3,\vec{k}_3)\phi(\omega_2,\vec{k}_2)\phi(\omega_1,\vec{k}_1)\rangle
\end{aligned}
\eeq
Since each factor of  $(\omega_i -k_i)$ will vanish in the limit $\omega_i \to k_i$, it is easy to see that only terms in the in-out correlator with three poles in on-shell limit will contribute to amplitude.  As result, only two terms from Equation~(\ref{eq:inin_fourier}) contribute to the amplitude
\beq\label{eq:LSZ_amp}
\begin{aligned}
\langle k_2,k_3|k_1\rangle 
&= i\mu (2\pi) \lim_{\omega_i \to k_i}\delta(\omega_2+\omega_3-\omega_1)
\Big(-\frac{\omega_3-k_3}{\omega_3-k_3}\frac{\omega_1-k_1}{\omega_3+\omega_2-k_1}\frac{\omega_2-k_2}{-k_1+k_2+k_3}\\
&\hskip 5cm +\frac{\omega_3-k_3}{\omega_3-k_3}\frac{\omega_2-k_2}{\omega_3-k_3+\omega_2-k_2}\frac{\omega_1-k_1}{-k_1+k_2+k_3}\Big)\\
&=-i\mu(2\pi)\delta(k_2+k_3-k_1)
\end{aligned}
\eeq
We see that the poles in $k_1-k_2-k_3$ are precisely those that give the $k_1 \to k_2 + k_3$ scattering amplitude.

The equal-time in-in and in-out correlation functions are the same, yet in both cases the poles responsible for a non-zero scattering amplitude vanish at equal time.  This is a reflection of the fact that no scattering process takes place in the vacuum: the only way to produce the scattering process requires one of the operators to be placed at $t\to -\infty$ and the other two taken at $t \to +\infty$.  The total energy pole that survives at equal time reflects only the $0\to 3$ or $3\to 0$ processes that are forbidden by energy conservation.

\subsubsection*{Classical Fluctuations}

Time-ordered (in-out) correlation functions are also calculable for classical statistics. Using the time-ordered Green's function, we can calculate the leading correction to $\varphi$ as
\beq
\phi^{(2)}(\k,t) =\frac{\mu}{2} \int dt' G_{\rm F}(k,t-t')\int \frac{d^3 p}{(2\pi)^3} \phi(\p,t')  \phi(\k-\p,t')  \ ,
\eeq
where we are using the Feynman propagator
\beq
 G_{\rm F}(k,t) = \frac{1}{2k} e^{-i k |t|} \ .
\eeq
We can then calculate the time-ordered correlator by substituting this expression and applying the classical (Gaussian) statistics,
\begin{align}\label{eq:classical_IO}
\langle T \phi(t, \k_1) \phi(t,\k_2) \phi(t,\k_3) \rangle_c &= - i \int^{\infty}_{-\infty} dt' \frac{\mu}{16 k_1 k_2 k_3} e^{- i k_1|t-t'|-i(k_2+k_3)(t-t') }+ {\rm permutations}\\
&= - \frac{\mu}{16 k_1 k_2 k_3 }\left(\frac{3}{  k_t} - \frac{1}{k_1-k_2-k_3}-\frac{1}{k_2-k_3-k_1}- \frac{1}{k_3-k_1-k_2} \right) \ . \nonumber
\end{align}
Like the quantum case, this is precisely the same as the in-in correlator and we see the appearance of poles at physical momenta.  

In order to understand these new poles, we first notice that the LSZ formula does not apply straightforwardly to our classical correlator.  Concretely, we recall that the relation between $\omega>0$ ($\omega <0$) and particles in the in-state (out-state) relied on the fact that the quantum vacuum is annihilated by the negative frequency mode, $a_\k |\Omega \rangle = 0$. To make sense of what LSZ would imply for our classical correlations, let us interpret the classical correlators as arising from highly occupied state, $|n\rangle$.  If we apply LSZ in this state, we get
\beq\label{eq:LSZ_class}
\begin{aligned}
\int \mathrm{d}^{4} z \frac{i e^{i q \cdot z}\left(-\square_{z}+m^{2}\right)}{(2 \pi)^{\frac{3}{2}} Z^{\frac{1}{2}}} & \left\langle n_{\text{out}}\left|\mathrm{T} \varphi\left(x_{1}\right) \ldots \varphi\left(x_{m}\right)
\varphi\left(z\right)\varphi\left(y_{1}\right) \ldots \varphi\left(y_{n}\right)\right| n_{\text{in}}\right\rangle \\
=&\sqrt{2\omega_p}\left\langle n_{\text{out}} \left|\mathrm{T} \varphi\left(x_{1}\right) \ldots \varphi\left(x_{m}\right)
\varphi\left(y_{1}\right) \ldots \varphi\left(y_{n}\right) a_p^\dagger \right| n_{\text{in}}\right\rangle\\
&-\sqrt{2\omega_p} \left\langle n_{\text{out}} \left| a_p^\dagger \mathrm{T} \varphi\left(x_{1}\right) \ldots \varphi\left(x_{m}\right)
\varphi\left(y_{1}\right) \ldots \varphi\left(y_{n}\right)  \right| n_{\text{in}} \right\rangle \ ,
\end{aligned}
\eeq
where we have labeled $|n_{\text{out}}\rangle$ and $|n_{\text{in}}\rangle$ to indicate that they are defined to $t = +\infty$ and $-\infty$ respectively.  We see that the LSZ formula does not isolate a particle in an in- or out-state, but instead isolated a particle in the in state minus a hole in the out state (or vice versa) for every field.  

A related consequence of this analogue of Equation~(\ref{eq:LSZ_class}) is that equal time correlators can now exhibit physical poles.  Specifically, we can use LSZ to calculate the S-matrix element,
\beq
\lim_{t\to +\infty} \left(\langle n_{\text{out}} | a_{k_2} a_{k_3}a^{\dagger}_{-k_1}\right) \, | n_{\text{in}}\rangle = \langle n_{k_2}+1, n_{k_3}+1, n_{k_1}-1 | n_{\text{in}}\rangle \propto {\cal A}_{1\to 2} \delta(\k_1 -\k_2 -\k_3) \ .
\eeq
This is just the amplitude for the decay of a particle with momentum $\k_1$ in the initial state to particles with momentum $\k_2$ and $\k_3$ in the final state.  Since all the operators acting on $\langle n|$ as $t\to +\infty$ do not vanish, there is no reason to expect the equal-time correlator to vanish either.  

The most straightforward consequence is that the poles at physical momenta seen in the equal-time in-out (and consequently in-in) correlators, Equation~(\ref{eq:classical_IO}), are the same poles responsible for the $1\to 2$ and $2\to1$ S-matrix elements determined by LSZ.  This can be seen by directly applying the (naive) LSZ formula to the unequal-time classical in-out correlator.  The result mirrors our quantum calculation.  The full expression for the non-equal time correlator is quite long, but the term responsible for the ``S-matrix" element is now, 
\begin{align}\label{eq:class_unequal}
\langle \varphi_{k_1}(t_1)\varphi_{k_2}(t_2)\varphi_{k_3}(t_3)\rangle^{\prime}_c \supset& \frac{\mu}{16k_1k_2k_3}\Bigg(\frac{\cos[k_2\left(t_1-t_2\right)+k_3\left(t_1-t_3\right)]}{k_1-k_2-k_3}
\\
&+\frac{\cos[k_1\left(t_1-t_2\right)+k_3\left(t_2-t_3\right)]}{-k_1+k_2+k_3}
+\frac{\cos\left[k_1\left(t_1-t_3\right)+k_2\left(t_3-t_2\right)\right]}{-k_1+k_2+k_3} \Bigg) \nonumber  \\
&\xrightarrow[]{t_1=t_2=t_3} \frac{\mu}{16k_1k_2k_3} \frac{1}{-k_1+k_2+k_3} \ .
\end{align}
In other words, if we were to repeat the LSZ procedure, as in Equation~(\ref{eq:LSZ_amp}), to the first two lines of~(\ref{eq:class_unequal}), we would recover a non-zero result.  The final line shows that this term survives the equal limit limit, in contrast to quantum case.  This provides a concrete demonstration that the physical poles in the in-in correlators can be interpreted as the decay of particles in the initial state, as was argued in~\cite{Green:2020whw} for inflationary correlators.  

\subsection{Resolving and Interpreting Poles at Physical Momenta}

The connection between in-in correlators and S-matrix elements provides some useful intuition.  However, if the in-in correlators represent a physical measurements, we would not expect them to have true poles at physical momenta, as the answers to physical questions are rarely infinite\footnote{It is also known that these divergences are not regulated by standard renormalization techniques, as was seen from studying the divergences of perturbation theory in non-Bunch Davies dS vacua~\cite{Banks:2002nv}. }.  In the
case of inflationary correlators, it was argued in~\cite{Green:2020whw} that the S-matrix elements that are responsible for the poles also cause the particles to decay, therefore introducing an effective width.  As a result, one might expect the physical poles to be replaced by a resonance, both in the inflationary context and in flat space.  While this resolution was seen in explicit examples~\cite{LopezNacir:2011kk,Turiaci:2013dka}, it is less clear that this the only resolution, particularly in cosmology.  In an expanding universe, the energy density blue shifts in the past and diverges as $t\to -\infty$.  As a result, cosmological correlators are also regulated by the finite duration of inflation~\cite{Holman:2007na,Agullo:2010ws,Ashoorioon:2010xg,Ganc:2011dy,Chialva:2011hc,Agullo:2012cs}.

Flat space provides a very useful testing ground for the regulation of these physical poles.  First, there is no analogue of the blueshift and the energy does not diverge at early times. As a result, the $t\to -\infty$ limit is not necessarily unphysical.  In addition, in flat space, decays can be forbidden by energy conservation\footnote{As energy is not conserved in cosmology, decays can always occur.} and thus can eliminate the role of a finite width.  This is easily achieved by considering massive scalars in place of our massless correlators.  Using this approach, we will see that it is not strictly cosmological expansion that is responsible for physical divergences as $t \to -\infty$.  

\subsubsection*{Massive Particles}

In flat space, the decays of particles are controlled by energy conservation.  The simplest way to test the connection between the physical poles and decay is to consider massive particles such that $E_k = \sqrt{k^2 +m^2}$ and the positive frequency mode functions become
\beq
\phi(\k, t) = \frac{1}{\sqrt{2 E_k}} e^{i E_k t} \ .
\eeq
With the modified mode function, we can calculate the equal time in-in bispectrum as before.  For quantum and classical statistics, one finds
\bea
\langle T \phi(0, \k_1) \phi(0,\k_2) \phi(0,\k_3) \rangle &=& - \frac{\mu}{4 E_1 E_2 E_3 (E_1+E_2+E_3) } , 
\eea
and
\beq
\begin{aligned}
\langle \phi(t, \k_1) \phi(t,\k_2) \phi(t,\k_3) \rangle_c =   \frac{\mu}{16 E_1 E_2 E_3} &\Bigg(\frac{3}{ E_t }  -\frac{1}{E_1-E_2-E_3}\\
&-\frac{1}{E_2-E_1-E_3}-\frac{1}{E_3-E_1-E_2} \Bigg) \ ,
\end{aligned}
\eeq
respectively.  While we see the poles when $E_1+E_2=E_3$, and permutations thereof, these poles cannot be reached at physical energies for the same kinematic reason that the lightest massive particle is stable.  As a result, we see the connection between the stability of these particles and the absence of poles at physical momentum. 

At first sight, this indeed suggests that the finite width of the particle is sufficient to avoid poles at physical momenta. At least for the three point function, eliminating the width also eliminated the pole. However, if we continue to higher point correlators, even stable particles can lead to poles as physical momenta.  We do not find such poles at four-points with a $\lambda  \phi^4$ interaction.  However, the classical five point correlator due to a contact interaction $\H_{\rm int} = \frac{1}{5!} \frac{\phi^5}{\Lambda}$ takes the form 
\beq
\langle \phi(t,\k_1) .. \phi(t,\k_5) \rangle_c = \frac{1}{256\Lambda E_1 E_2 E_3 E_4 E_5} \bigg( \frac{5}{E_{\rm tot}} + \sum_{i} \frac{3}{(E_{\rm tot}-2 E_i)} + \sum_{i\neq j} \frac{1}{E_{\rm tot} - 2 E_i -2 E_j} \bigg)
\eeq
where $E_{\rm tot}= \sum_i E_i$. The final term contains a pole that is consistent with the allowed kinematic region of $2\to 3$ scattering.  In this sense, we can see that nothing prevents us from reaching this pole for physical momentum. In addition, since the particles don't decay, there is no finite width that needs to be included in this calculation.  Finally, there is no analogue of the blue-shifting of energies at early times that demands that we regulate the early time limit of this calculation.  Clearly we need another physical interpretation for how this pole arises.  

\subsubsection*{Finite Time of Interactions}\label{sec:finite_time}

The origin of the physical poles in the classical case can be understood from the integral expression for the correlator,
\beq
\langle \phi(t,\k_1) .. \phi(t,\k_5) \rangle_c = \frac{1}{16 \Lambda E_1.. E_5} \sum_i \int_{-\infty}^t dt' \sin(E_i (t-t')) \prod_{j\neq i} \cos(E_j (t-t'))
\eeq
When we sit on a pole where $E_{\rm tot} - 2 E_i -2 E_j$, there is a non-oscillatory contribution to the integrand such that the integral diverges at $t \to -\infty$.  This is a reflection of the fact that there is now an on-shell process that changes the classical distribution.  Specifically, our Gaussian state $|n\rangle$ is not a stationary configuration in the presence of these interactions.  Instead, the particles can now scatter, exchanging energy, momentum, and even particle number.  Given an infinite amount of time to interact, we should expect the final distribution of particles to be a stationary configuration, e.g.~a thermal distribution.  Stationary states do not normally exhibit the long range correlations of the quantum vacuum fluctuations\footnote{We are not claiming long range correlations are impossible in general, but the consistent appearance of these poles in perturbation theory suggests that it does not arise from a generic local Hamiltonian near a Gaussian fixed point.}.

We should therefore create the initial Gaussian state at a finite time in the past, $t_i < t$, such that the distribution at time $t$ is only weakly non-Gaussian.  For simplicity, we return to the bispectrum
\begin{align}
\langle \phi(t, \k_1) \phi(t,\k_2) \phi(t,\k_3) \rangle_c =&  \frac{\mu}{16 k_1 k_2 k_3} \Bigg(\frac{3\left(1- \cos(k_t \Delta t)\right)}{ k_t } \\
& +\bigg( \frac{1- \cos\left((k_1-k_2-k_3) \Delta t\right)}{k_1-k_2-k_3}+{\rm permutations} \bigg) \Bigg)
\end{align}
where $\Delta t = t-t_i$.  Unlike a mass or width which moves the poles to complex momenta, now we see that there are no poles at all.  Instead, when we take $k_1 \to k_2+k_3$
\beq
 \frac{\mu}{16 k_1 k_2 k_3} \frac{1- \cos\left((k_1-k_2-k_3) \Delta t\right)}{k_1-k_2-k_3}  \to \frac{\mu}{32 (k_1+ k_2)k_2 k_3} (\Delta t)^2 (k_1-k_2-k_3) \ ,
\eeq
which vanishes as $k_1-k_2-k_3 \to 0$.  This correlator gets its largest contribution when $(k_1-k_2-k_3) \approx \Delta t^{-1}$ and is enhanced relative to the total energy pole by a factor $\Delta t k_t$.  As a result, the signal in the folded configurations will still dominate over the equilateral~\cite{Babich:2004gb}, much like the cosmological setting~\cite{Green:2020whw},  .  This is consistent with more general expectation about the perturbative structure of cosmological correlators.  On general grounds, even the total energy pole is expected to vanish for cosmological correlators in a UV complete theory~\cite{Maldacena:2015iua,Arkani-Hamed:2015bza}.  Yet, in perturbation theory the poles accurately capture the observable signals~\cite{Smith:2006ud}.  

\section{Particle Detectors and the In-In Formalism}\label{sec:detector}

The structure of in-in correlation in flat space is largely the same as in (quasi-) de Sitter space.  However, without cosmological particle production, we don't have an obvious interpretation of the correlator in terms of some physical process.  One might worry that this is some mathematical devise that lacks a physics reality outside of cosmology.  In this section, we will show how the in-in correlator arises in physical models of particle detection.  This will allow us to give a clear physical meaning to the flat space correlators and their poles.

\subsection{Unruh -- de Witt Detectors}

We will first review the Unruh-de Witt (UdW) model for particle detection.  The central idea is that we have a single qubit, that registers whether or not there was a particle in some localized region of space-time.  To do so, we place it in the zero-state initially ($\zk$).  We then couple this qubit to our field, $\phi(\x, t)$, for some finite amount of time and inside some localized region of space.  After we turn off the coupling to $\phi$,  we should have a non-zero probability of finding our qubit in the one-state, $\ok$, if there was a particle (or were particles) in the detector while it was on.

We implement this model by introducing an interaction Hamiltonian that couples our qubit to $\phi$.  Following \cite{Unruh:1983ms}, the detector is described by a Hamiltonian
\beq
H_{\rm D}=\lambda \, \epsilon(t) \int d^3 x \phi(\x,t)\left[\psi(\x) \hs+\psi^{*}(\x) \hs^{\dagger}\right] \ ,
\eeq
where $\lambda$ is the coupling constant, $\epsilon(t)$ is a function that defines how we turn on/off the detector, and $\psi(\x)$ is a function that defines the spatial resolution of our detector such that $\psi(x)$ vanishes outside the detector. The detector state is defined by
\beq
\hs \zk =\hs^{\dagger}\ok =0 \qquad \hs^{\dagger} \zk = \ok \qquad \hs \ok= \zk  \ ,
\eeq
and the free field is again given by
\beq
\phi(\x,t) = \int \frac{d^3 k}{(2\pi)^3} \frac{1}{\sqrt{2 k}} e^{i \k\cdot \x} [ a^\dagger_{-\k}  e^{i k t} + a_{\k}  e^{-i k t}  ] \ .
\eeq
Even though we are giving a quantum description of the detector, the fluctuations of field $\phi$ may be quantum or classical.

Let us check that, at leading order in $\lambda$, this detector works as promised.  We will put the scalar field into a single particle state
\beq\label{eq:prob_detect}
|\phi_1(\y,t_0) \rangle = \phi(\y) |0\rangle = \int \frac{d^3 p}{(2\pi)^3} e^{-i\p\cdot \y} f(p, t_0) a^{\dagger}_\p | 0\rangle \ ,
\eeq
where $f(p,t) = e^{i p t} /\sqrt{2p}$ as before.  If we now turn on the detector interaction, the probability for finding an excited detector at leading order,  ${\cal O}(\lambda^2)$, is 
\beq
P_{1} = |{\cal A}_{1;1\to 0}|^2 +|{\cal A}_{1;1\to 2}|^2
\eeq
where 
\bea
{\cal A}_{1;1\to 0} &=& \langle 0|  \lambda \int dt \epsilon(t)  \int d^3 x \psi^{*}(\x) \int \frac{d^3 p}{(2\pi)^3} e^{i \p \cdot(\x-\y)} f^*(p,t)  f(p,t_0) |0\rangle  \\
&=&  \lambda \int dt \epsilon(t)  \int d^3 x \psi^{*}(\x) G_{\rm F}(\x, t; \y, t_0)
\eea
and
\bea
{\cal A}_{1;1\to 2} &=&  \langle 2|  \lambda \int dt \epsilon(t)  \int d^3 x \psi^{*}(\x) \int \frac{d^3 p}{(2\pi)^3} e^{i \p \cdot(\x-\y)} \phi(-\p,t) \phi_0(\p) a^{\dagger}_p |0\rangle  \\
&=&  \lambda \int dt \epsilon(t)  \int d^3 x \psi^{*}(\x)\langle 2 | \phi(x,t) \phi(y,t_0) |0 \rangle \ .
\eea
The first term, ${\cal A}_{1;1\to 0}$, is the amplitude that the particle at $\y$ and time $t_0$ is absorbed at time $t$ in the detector located at $\x$.  This term captures the physics of interest, namely the detection of the particle that we put in the initial state.  The second term, ${\cal A}_{1;1\to 2}$, the probability that the detector creates an anti-particle\footnote{The real scalar $\phi$ is its own anti-particle, but this distinction is helpful for againing intuition.  See Appendix~\ref{app:nonlocal} for more details. } from the vacuum while registering a particle in the detector.  This contribution is not the detection of our initial particle, but is the detection of a particle created from the vacuum by the detector.

One important aspect of the UdW detector is that it shows concretely that the act of measuring the field $\phi$ changes the state of the system.  In particular, if we try to measure a particle in the vacuum (of the free theory), a non-zero amplitude for exciting the detector (at order $\lambda$) requires the creation of an anti-particle.  We can see this, in analogy with Equation~(\ref{eq:prob_detect}), by projecting onto a 1-particle final state, 
\beq
{\cal A}_{1; 0 \to 1} =  \lambda \int dt \epsilon(t)  \int d^3 x \psi(\x)  \langle 1 | \phi(x,t) |0 \rangle  \ .
\eeq
We can interpret this as follows: the act of performing the measurement absorbs a particle from a particle-anti-particle pair, creating an outgoing anti-particle of equal and opposite momentum in the process.  Importantly, it is the act of localizing the measurement in time and space that provides the energy and momentum needed to excite the vacuum.  This explanation coincides with our intuition from the uncertainty principle.  

To confirm the interpretation, let us consider what happens as we change the function $\epsilon(t)$ to be less localized in time, thus corresponding to smaller energies by the uncertainty principle.  If we create the one-particle state at a time $t_0 < 0$ and measure at $t\approx 0$ with
\beq
\epsilon(t) = \frac{1}{\sqrt{2\pi} \sigma_t} e^{-t^2/(2 \sigma_t^2)} \qquad \psi(\x) = \delta(\x) \ ,
\eeq
then the amplitude becomes
\beq
{\cal A}_{1; 0 \to 1} =  \lambda \int dt \epsilon(t)   \frac{1}{2 E_\k} e^{i E_\k t} = \frac{\lambda}{2 E_k} \, e^{-E_\k^2 \sigma_t^2/2} \ .
\eeq
This is, again, nothing more than the uncertainty principle, as our resolution in energy is inversely proportional to our resolution in time, $\sigma_E \propto \sigma^{-1}_t$.  Given that we start in the vacuum (zero energy), we must have a large uncertainty in energy to create particles.  For example, if we were to work with massive particles such that $E_\k \geq m $ for all $\k$, the probability of finding a particle in the vacuum is exponentially suppressed unless $\sigma_E \gg m$ or, equivalently, $\sigma_t \ll 1/m$.

\subsection{Particle Detection and Cosmological Correlators}

Now suppose we want to measure correlations of these vacuum fluctuations using our particle detector\footnote{Here we are using particle detectors in flat space, but it is interesting to also consider these detectors as a probe of inflation directly~\cite{Kaplanek:2019vzj}.}.  We can imagine placing $N$ UdW detectors at distinct points in space $\x_i$ such that the detectors do not overlap, $\psi(\x-\x_i)\psi(\x-\x_j) = 0$ for $i\neq j$.  Furthermore, we will assume that $\epsilon(t)$ is sufficiently localized in time such that the detectors are all space-like separated when they are on. We denote $|\Omega \rangle$ as the interacting vacuum of $\phi$ when $\lambda = 0$, so that the amplitude for all $N$ detectors to be in the $\ok$ state {\it and} for $\phi$ to remain in the vacuum is 
\beq
{\cal A}_{N;\Omega} =\left( \prod_i  \int dt'_i d^3 x'_i \lambda \epsilon(t-t'_i) \psi(\x_i -\x_i') \right) \langle \Omega | \phi(t'_1,\x'_1) .. \phi(t_N',\x'_N) |\Omega \rangle + {\rm local} \ .
\eeq
We see the this amplitude is proportional to the in-in correlator, convolved with the detector.  Because the detectors are spacelike separated, from Equation~(\ref{eq:inin}) we see that any additional terms associated with the commutator of the interaction Hamiltonian, $H_{\rm int}$, with the detector Hamiltonian, $H_{\rm D}$, will give purely local terms (i.e.~these contributions are equivalent to field redefinitions of $\phi$ and do not produce a total energy pole).  We can write this in terms of the in-in correlation function in fourier space as
\beq
{\cal A}_{N; \Omega} =\left(\lambda^N \prod_i  \int dt'_i d^3 k_i  e^{-i \k_i \cdot (\x_i-\x_i')}  \tilde \epsilon(t-t'_i) \tilde \psi(\k_i) \right) \langle \Omega | \phi(t'_1,\k_1) .. \phi(t_N',\k_N) |\Omega \rangle \eeq
where we used the notation where $\tilde \psi(\k)$ is the fourier transforms of $\psi(\x)$.   We see that the fourier modes that contribute to this amplitude are only those that appear in the detector itself via $\tilde \psi(k)$.

In the case where $\epsilon(t) \approx \delta(t)$ and $\psi(\x) \approx \delta(\x)$ with $t'_i \to t$, the amplitude becomes proportional to the equal time in-in correlator in position space:
\beq
{\cal A}_{N;\Omega} \to \lambda^N  \langle \Omega | \phi(t,\x_1) .. \phi(t,\x_N) |\Omega \rangle \ .
\eeq
 For quantum vacuum fluctuations, the amplitude for three coincident particles becomes 
\bea
{\cal A}_{3;\Omega} &=& \lambda^3 \langle  \phi(\x_1,t) \phi(\x_2,t) \phi(\x_3,t) \rangle \label{eq:3_detectors}\\ &=& -\lambda^3\frac{\mu}{4} \left[ \prod_i \int \frac{d^3 k_i}{(2\pi)^3} e^{i \k_i \cdot \x_i} \right] \frac{1}{(k_1+k_2+k_3)}  (2\pi)^3 \delta(\k_1+\k_2 +\k_3)\\
&=&-\lambda^3\frac{\mu}{4} \int \frac{d^3 k_1}{(2\pi)^3} e^{i \k_1\cdot \x_{13} }\int \frac{d^3 k_2} {(2\pi)^3}e^{i \k_2\cdot \x_{23} } \frac{1}{(k_1+k_2+k_3)k_1 k_2 k_3 } \ ,
\eea
where we defined $\x_{ij} \equiv \vec x_i - \vec x_j$.  In order to make the $\k_i $-integrals manageable, let us assume $x_2$ is far from $x_{1}$ and $x_3$ so that $x_{13} \ll x_{23}$.  We can then expand in $k_1 \approx k_3 \gg k_2$ to find
\begin{align}
\langle  \phi(\x_1,t) \phi(\x_2,t) \phi(\x_3,t) \rangle &\approx -\frac{\mu}{4} \int \frac{d^3 k_1}{(2\pi)^3} e^{i \k_1\cdot \x_{13} }\int \frac{d^3 k_2} {(2\pi)^3}e^{i \k_2\cdot \x_{23} } \frac{1}{2k_1^3 k_2} \\
&\approx \frac{\mu }{32 \pi^4} \frac{\log x_{13}}{x_{23}^2} \ . \label{eq:quantum_pos}
\end{align}
For comparison, the two point function of the massless field $\phi$ will fall off like $1/x^2$ ($\phi$ is dimension one in 3+1 dimensions).  As we separate the detectors, it is important that the contribution to the amplitude will decay with the same power of the distance as the it would in the free theory.

In contrast, let us see what happens to the detector in the presence of classical fluctuations.  We again take the in-in three point function and we will focus on the contribution from a physical pole (the total energy pole we give the same result as the quantum theory), 
\begin{align}
\langle  \phi(\x_1,t) \phi(\x_2,t) \phi(\x_3,t) \rangle_c \supset& \frac{\mu}{16} \left[ \prod_i \int \frac{d^3 k_i}{(2\pi)^3} e^{i \k_i \cdot \x_i} \right] \frac{1}{k_1 k_2 k_3 (k_1+k_2-k_3)}  (2\pi)^3 \delta(\k_1+\k_2 + \k_3) \nonumber \\
\supset&\frac{\mu}{16} \int \frac{d^3 k_1}{(2\pi)^3} e^{i \k_1\cdot \x_{13} }\int \frac{d^3 k_2} {(2\pi)^3}e^{i \k_2\cdot \x_{23} }\frac{1}{k_1 k_2 k_3} \frac{1}{k_1+k_2-k_3}  \ ,
\end{align}
where in the second line $k_3 = |\k_1+\k_2|$.  Again we will assume the point $x_2$ is far from $x_{1}$ and $x_3$.  As such we consider the limit $x_{13} \ll x_{23}$  by expanding in $k_1 \approx k_3 \gg k_2$.
\bea
\langle  \phi(\x_1,t) \phi(\x_2,t) \phi(\x_3,t) \rangle_c &\approx &\frac{\mu}{16} \int \frac{d^3 k_1}{(2\pi)^3} e^{i \k_1\cdot \x_{13} }\int \frac{d^3 k_2} {(2\pi)^3}e^{i \k_2\cdot \x_{23} } \frac{1}{k_1^2 k_2^2(1-\cos \theta)} \\
 &\approx & \frac{\mu}{64 \pi^4} \frac{1}{x_{13} x_{23}}\left(- \log \left( \frac{\theta_{\rm min}^2}{2} \right)  \, f(\hat x_{23} \cdot \hat x_{13}) \right) \label{eq:classical_position} \ ,
\eea
where $\theta_{\rm min}$ is the minimum angle between $\k_1$ and $\k_2$ and $f(\hat x_{23} \cdot \hat x_{13})$ is a function of the angle between $\vec x_{13}$ and $\vec x_{23}$ with $f(1) = 1$.  This formula is noteworthy for two reasons: first, it has a pole as $x_{13} \to 0$ which enhances the size of the non-Gaussian signal.  Second, it falls off more slowly than the Gaussian two-point function and will therefore give the dominant contribution to any long distance correlations.

\subsection{Classical Interpretation}

We have seen how the detector model provides a physical interpretation of the quantum in-in correlators.  For quantum fluctuations, we can naturally understand the role of the detector in exciting the vacuum and giving rise to particles.  The uncertainty principle tells us that localizing a measurement in time will mean that we are no longer in an energy eigenstate (in this case, the vacuum).  This is an inevitable feature of a quantum-measurement involving an operator that doesn't commute with the Hamiltonian.  

Stated this way, it is the interpretation of the classical measurement that requires explanation.  Classical measurements do not have to disturb the state and therefore the response of our detector should be a fundamental property of the classical system.  On the other hand, our derivation of the amplitude for exciting the detector did not assume the in-in correlator was calculated in the vacuum and would be equally applicable in the classical limit using
\beq\label{eq:classical_inin}
\langle \phi(\k_1) .. \phi(\k_N) \rangle_{c}   = \lim_{n\to \infty} 2 {\rm Im} \langle n | \phi(\k_1) .. \phi(\k_N)  \int^t dt' H_{\rm int} (t') | n  \rangle \ .
\eeq
For this to be consistent with something classical, it must be how we interpret the state of the detector that has to change.  For states close to the quantum vacuum, we demonstrate that the UdW detector is designed to be excited when a particle is present in the initial state.  In the classical case, we have lots of particles in the initial state, even in the absence of fluctuations, and therefore the response of the detector to this state requires more care.  

Consider what happens as we evolve the state $|n \rangle$ in the presence of the detector Hamiltonian, $H_D$.  Working to linear order in $\lambda$, we get at time $t$,
\begin{align}
\zk \, |n\rangle  \to& \zk \, |n\rangle + i \lambda \ok \int d^3 x \psi(x) \hat \phi(\x,t) |n \rangle \\
=& \zk \, |n\rangle \\
&+ i \lambda \ok \int \frac{d^3 k}{(2\pi)^3} \frac{\psi(-\k)}{\sqrt{2 E_k}} \left( e^{i k t} \sqrt{n_{-\k}+1} |n_{-\k}+1\rangle |\hat n;  -\k \rangle +e^{-i k t} \sqrt{n_{\k}}  |n_{\k}-1\rangle |\hat n;  \k \rangle \right) \nonumber \ .
\end{align}
The second term shows that the detector will register a ``particle" for both an increase and decrease in the total number of particles.  In addition, we notice that in the limit $n_k \to \infty$, this is approximately
\begin{align}
\lim_{n \to \infty} \zk \, |n\rangle 
\to& \zk \, |n\rangle+ i \lambda \ok \int \frac{d^3 k}{(2\pi)^3} \frac{\psi(-\k)}{\sqrt{2 E_k}} \sqrt{n_k} \cos(k t) |n\rangle  \ .
\end{align}
Concretely, our state is responding to the classical (real) oscillations of $\phi$ around the mean density.

In the free theory, there was no problem working in this very carefully defined excited state, $|n\rangle$. Including interactions not only produces higher-point correlations, it also takes the state away from the static initial state we took in the Gaussian theory.  Since the detectors are essentially registering changes in the state away from the Gaussian initial state, this classical evolution alone should be sufficient to excite the detectors.  Evolving the state according the interacting Hamiltonian, we find
\begin{align}
|n\rangle \to& |n\rangle + \int dt' H_{\rm int} (t') | n  \rangle \\
=&|n\rangle + \frac{\mu}{3!} \int dt' \int d^3 x \phi^3(\x,t') |n\rangle \\
=& |n\rangle + \frac{\mu}{3!} \int dt'  \bigg( \prod_i \int \frac{d^3 k_i}{(2\pi)^3}\frac{1}{\sqrt{2 E_{k_i}}} \Big( e^{i k_i t'} \sqrt{n_{-\k_i}+1} |n_{-\k_i}+1\rangle |n_{\k_i}\rangle
\\
&  \qquad +e^{-i k_i t'} \sqrt{n_{\k_i}}  |n_{\k_i}-1\rangle|n_{-\k_i}\rangle \Big) \bigg)\delta^{(3)}(\sum_i \k_i) |\hat n; \k_1,\k_2,\k_3, -\k_1,-\k_2,-\k_3\rangle \ . \label{eq:classical_evolution}
\end{align}
We see that the time evolution of state changes the number (density) of particles of different momenta.  This change to the density is precisely of the form that our UdW detector will register as a ``particle".  Classically, the detector is not responsible for exciting the fluctuations, those were already created by the classical-time evolution of $\phi$.

An additional confusion with this description is the meaning of the total energy pole.  The poles at physical momenta capture the decay of the particles in the initial state which leads to density fluctuations by changing the number densities of particles with different momenta.  The total energy pole does not have such a simple description classically.  In the quantum theory, we interpreted this pole as a violation of energy conservation, via the uncertainty principle, which would be forbidden in a classical theory.  However, the quantum interpretation was needed because the quantum vacuum is an energy eigenstate of the full interacting Hamiltonian. In contrast, what we see in Equation~(\ref{eq:classical_evolution}) is that the state $|n \rangle$ is not a stationary state of the classical Hamiltonian. In the presence of the interaction, the system wants to evolve towards equilibrium as explained in Section~\ref{sec:finite_time}.  The total energy pole is therefore not a violation of energy conservation but a reflection that our classical probabilistic system included states of different energies initially. 

\section{Detector Entanglement}\label{sec:entanglement}

Entanglement has become an increasingly valuable probe of fundamental physics.  It can reveal the structure of quantum field theories and states of matter in flat space~\cite{Casini:2022rlv}. In curved space times, it is central to our understanding of black holes~\cite{Page:1993wv,Almheiri:2020cfm,Bousso:2022ntt} and particle production~\cite{Maldacena:2012xp}. Entanglement is even thought to encode the causal structure of spacetime itself~\cite{VanRaamsdonk:2010pw,Maldacena:2013xja,Rangamani:2016dms}.   

In the quantum field theory, the entanglement entropy in vacuum is expected to follow an area law~\cite{Srednicki:1993im,Eisert:2008ur}, while it should scale as volume in a generic excited state.  This offers a different starting point for understanding the nature of the quantum vacuum than the structure of poles in cosmological correlators.  In this section, we will explore to what degree these properties are related. 

In order to make the comparison, we will consider $N$ UdW-detectors and some free field $\phi$ such that the state of the detectors is
\beq\label{eq:N_detectors}
\begin{aligned}
\left|\Psi_{\phi,{\rm UdW}} \right\rangle &= (\mathbf{1}-C)|0_{\{i\}}\rangle -i \sum_j \Phi_{j}|\Omega\rangle |1_j,  0_{\{i, \hat j \}} \rangle-\sum_{j, k}\Phi_{j} \Phi_{k}|\Omega\rangle |1_j 1_k, 0_{\{i, \hat j, \hat k\}}  \rangle +O\left(\lambda^{3}\right) \ ,
\end{aligned}
\eeq
where $\Phi_{i}=\lambda \int d t \epsilon_{i}(t)\int d^{3} x \psi_{i}(\vec{x}) \phi(\vec{x}, t)$, $C$ is a normalization constant, and the detector states are defined by 
\beq
|1_j,  0_{\{i, \hat j \}} \rangle \equiv  \hs^{\dagger}_j | 0 \rangle_{\boldsymbol{s}_j} \bigotimes_{i \neq j} | 0 \rangle_{\boldsymbol{s}_i} \qquad |1_j 1_k, 0_{\{i, \hat j, \hat k\}}\rangle  \equiv  \hs^{\dagger}_j | 0 \rangle_{\boldsymbol{s}_k}   \otimes  \hs^{\dagger}_k | 0 \rangle_{\boldsymbol{s}_k} \bigotimes_{i \neq j,k} | 0 \rangle_{\boldsymbol{s}_i} \ .
\eeq
This setup is the $N$-particle generalization of the detector configuration described in~\cite{Reznik:2002fz,Reznik:2003mnx}.

We will be interested in understanding the entanglement of the individual detectors as a probe of the state of $\phi$.  As the detectors already encode the cosmological correlators, this offers a simple approach to connecting these correlations with the entanglement.  Our approach will be qualitatively similar {\it entanglement harvesting}~\cite{Salton:2014jaa,Hummer:2015xaa}.  However, given our cosmological motivations, we will not be interested in whether the detectors themselves are in a uniquely quantum state (cosmological observations are always classical after all).  Instead we want to use it as a probe of the quantum nature of $\phi$ itself.  

We will assume that we know that $\phi$ is in the interacting vacuum.  This is a useful assumption as we can project the state in Equation~(\ref{eq:N_detectors}) onto the vacuum, so that the state of the detector becomes
\begin{align}
\left|\Psi_{\rm UdW}\right\rangle \equiv \langle \Omega \left|\Psi_{\phi,{\rm UdW}} \right\rangle &= (\mathbf{1}-C)|0_{\{i\}} \rangle-\sum_{j,k} \langle \Omega | \Phi_{i}^{+} \Phi_{j}^{+}|\Omega \rangle |1_j 1_k, 0_{\{i, \hat j, \hat k\}}  \rangle +O\left(\lambda^{3}\right) \ .
\end{align}
  As a theoretical tool, this projection has the advantage that it leaves the detectors in a pure quantum state. In other words, we have projected onto the product state so that the density matrix takes the form $\rho_{\phi,{\rm UdW}}=\rho_{{\rm UdW}}\bigotimes | \Omega\rangle\langle  \Omega|$.  As a result, the entanglement between the detector and the field degrees of freedom has been set to zero, $S_{\text{detector}}= S_{\phi}=0$.  Note that this procedure does not represent a true measurement, as the purpose of the detectors is to measure the state of the field $\phi$, which we presumably to not know directly.

\subsection{Position Space Entanglement}

First, we will assume that each detector is localized at a point in space-time so that the detector is described by $\psi_i(\x) \approx \delta(\x-\z_i)$ for some point $\z_i$ and $\epsilon(t) \approx \delta(t)$.  Note that with these choices, $\lambda$ has units of distance.  We now want to use the entanglement between the UdW detectors as a proxy for the entanglement in the underlying state of $\phi$.  Having localized the detectors in space, it is natural to consider entanglement between the detectors localized in two regions $A$ and $B$.  

For simplicity, let us take the region $A$ as a sphere of radius $R$ and $B = \overline{A}$ is the region outside the sphere.  Now we want to define the state of the detectors located inside $A$ and $B$, $\vec x_i \in A$ and $\vec y_j \in B$, as follows:
\begin{align}
|0,0\rangle&\equiv |0_{\{ i \}} \rangle  \\
|\{ \x_n \} ,0\rangle &\equiv  |1_{\x_1} ..1_{\x_n}, 0_{\{i, \hat x_{1},..,\hat x_n \}} \rangle  \\
 |0,\{ \y_N \} \rangle &\equiv  |1_{\y_1} ..1_{\y_N}, 0_{\{i, \hat y_{1},..,\hat y_N\}} \rangle \\
  |\{ \x_n \},\{ \y_N \}\rangle &\equiv  |1_{\x_1} ..1_{\x_n},1_{\y_1} ..1_{\y_N}, 0_{\{i, \hat x_{1},..,\hat x_{n},\hat y_1,..,\hat y_{N} \}} \rangle
\end{align}
so that
\beq\label{eq:detector_state_position}
\begin{aligned}
\left|\Psi_{\rm UdW}\right\rangle=&|0,0\rangle+\sum_{\{x_n\} \neq \varnothing} a_{\{\x_n\}}|\{ 
\x_n \}, 0\rangle+\sum_{\{\y_N\} \neq \varnothing} b_{\{\y_N\}}|0, \{\y_N\} \rangle\\
&+\sum_{\{\x_n\},\{\y_N\}  \neq \varnothing} c_{\{\x_n\}, \{\y_N\}}|\{\x_n\}, \{\y_N\}\rangle \ .
\end{aligned}
\eeq
The entanglement between $A$ and $B$ can be determined from reduced density matrix of the detectors in $A$ (inside the sphere) tracing over the detectors in $B$ (outside the sphere).  The contributions from $a_{\{\x_n\}}$ and $ b_{\{\y_N\}}$ can be removed by a change of basis (as they do not entangle detectors in $A$ with detectors in $B$).  The contribution to a non-trivial density matrix comes instead from
\beq
\rho_{\{\x_n\},\{\x_m\}} = \sum_{\{\y_N\}}  c_{\{\x_n\},\{\y_N\}} c^{\dagger}_{\{\x_m\},\{\y_N\}}|\{\x_n\} \rangle \langle \{\x_m\}|  \ .
\eeq
In the free theory, these coefficients are determined by Wick contractions alone, 
\beq
c_{\{\x_n\} ,\{ \y_N \}} = \prod_{i,j} \langle \Phi_{\x_i} \Phi_{\y_j} \rangle + {\rm permutations}  \ .
\eeq
For a massive theory, these two point functions are exponentially suppressed and therefore
\beq
c_{\x_i ,\y_j } \propto e^{-m |\x_i-\y_j|}  \ .
\eeq
Any non-trivial entanglement between the detectors in $A$ and $B$ will therefore be exponentially suppressed as well.

For massless fields, correlations scale like powers for the distance and thus are more entangled than the massive case. Expanding in $\lambda$, the leading contribution is from 
\beq\label{eq:free_2pt}
c_{\x_i,\y_j} =\langle \Phi_{\x_i} \Phi_{\y_j} \rangle = \frac{\lambda^2}{|\vec x_i-\vec y_j|^2} \ ,
\eeq
where we assumed we are in $3+1$ dimensions.  The key question we want to understand is how $B$ and $A$ are entangled. We are particularly interested in how the reduced density matrix of the detectors within $A$ depends on their proximity to the boundary with $B$. In massive QFT, entanglement is short ranged and is responsible for the famous area-law in the quantum vacuum. We now want to understand what happens for the range of entanglement of our detectors. For massive fields, the exponential decay of the two point correlators ensures the entanglement is short range.

To understand the range of entanglement of the detectors for massless $\phi$, we will consider a small number of localized detectors with $A$ entangled with detectors in $B$.  The simplest such possibility is two detectors, one in $A$ and one in $B$.  The state $| \Psi_{\rm UdW} \rangle$ of these detectors gives us a density matrix at ${\cal O}(\lambda^4)$
\beq
\rho^{(AB)}=\left(\begin{array}{cccc}
1-c_{\x,\y}^2 & 0 & 0 &  -c_{\x,\y} \\
0 &0 & 0 & 0 \\
0 & 0 & 0 & 0  \\
-c_{\x,\y} & 0 &0 & c_{\x,\y}^2
\end{array}\right) \ ,
\eeq
where the matrix indices are $|0_\x 0_\y\rangle$, $|1_\x 0_\y\rangle$, $|0_\x 1_\y\rangle$ and $|1_\x 1_\y\rangle$.  This is, of course, the density matrix of a pure state\footnote{Notice this density matrix differs from those in similar two-detector models as in~\cite{Reznik:2002fz,Reznik:2003mnx}.  The difference arises because we have projected onto the ground state of $\phi$ to arrive at a pure state, rather than tracing over $\phi$, which results in a mixed state.}.  Tracing over $B$ produces the reduced density matrix for the detector in $A$,
\beq
\rho^{(A)}= {\rm Tr}_B \rho^{(AB)} =\left(\begin{array}{cc}
1-c_{\x,\y}^2 & 0  \\
0 &c_{\x,\y}^2 
\end{array}\right) \ ,
\eeq
and entanglement entropy
\beq
S^A_{\rm ent}= - {\rm Tr} \rho^{(A)} \log \rho^{(A)} = - c_{\x,\y}^2 (\log c_{\x,\y}^2 - 1) +{\cal O}(\lambda^6) \ .
\eeq
We see that in the limit $y \to \infty$, holding $\x$ fixed, $S^A_{\rm ent} \propto y^{-4} \log y \to 0$.

Now let us consider the generalization of this case where we take a single detector in $A$ at location $\x$ with a large number of detectors in $B$, $N_B$, that are uniformly distributed in space.  The entanglement between any the detector in $A$ and any detector $B$ is the same as our above example, such that the reduced density matrix is again diagonal with $\rho_{00} = 1-\rho_{11}$ and 
\beq
\rho_{11} \approx \lambda^4 \sum_{\y_j} \frac{1}{|\x - \y_j|^4}  \ .
\eeq
First let us place the detector in $A$ near then center of the sphere so that $x \ll R$. Using the large $N_B$ limit to replace the sum by an integral, the leading contribution to the reduced density matrix becomes 
\beq\label{eq:N_B}
\rho_{11}  \approx  \frac{4\pi \lambda^4 N_B}{V} \int_R^\infty r^2 dr \frac{1}{r^4} = \frac{4\pi \lambda^4}{R} n_B \ ,
\eeq
where $n_B = N_B/V$ is the number density of detectors in $B$.  In contrast, a point near the boundary gets a much larger contributions.  E.g. if $x \approx R$ we can write
\beq
\rho_{11}   \approx \frac{2 \pi \lambda^4 N_B}{V} \int^\infty_{1/\Lambda} r^2 dr \frac{1}{r^4} = 2\pi \lambda^4 \Lambda n_B \ ,
\eeq
where $\Lambda^{-1}$ is the smallest distance between a detector in $B$ and $\x$.  So far, the entanglement of a few detectors inside $A$, for a free theory of $\phi$, are consistent with the intuition that entanglement mostly arises from point near the boundary of $A$, or that entanglement in the vacuum is short ranged\footnote{We  can also repeat this argument for generalized free field (e.g.~\cite{Dymarsky:2014zja}) by substituting $|\x-\y|^{-2} \to |\x-\y|^{-2\Delta}$ in Equation~(\ref{eq:free_2pt}),  where $\Delta$ is the scaling dimension of the operator.  It is interesting to note that the unitarity bound $\Delta\geq 1$ ensures that entanglement of the detectors is short ranged.}. 

 \begin{figure}[h!]
\centering
\includegraphics[width=3.5in]{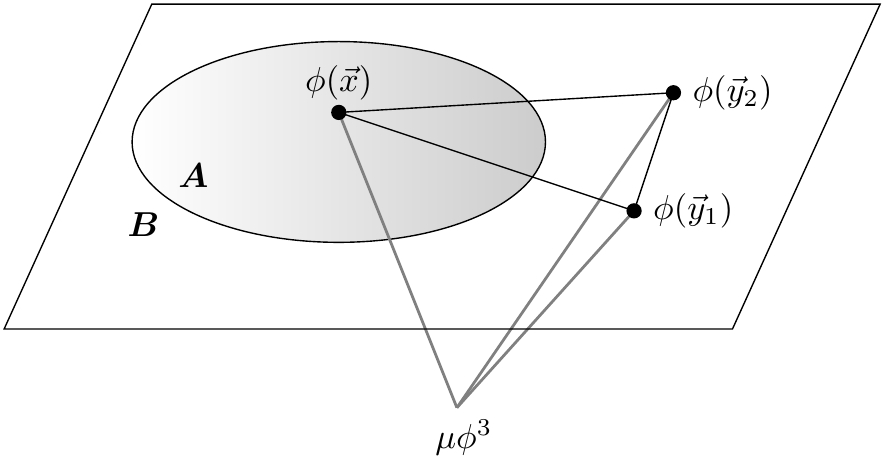}
\caption{Entanglement between detector in region $A$, located at $\x$, and two detectors in $B$, located at $\y_1$ and $\y_2$, produced by the three point interaction $\H_{\rm int} = \mu \phi^3$.  Time runs in the vertical direction such that the correlation between detectors on the equal time surface is due to interactions in the past. }
\label{fig:entangle}
\end{figure}

Now we would like to understand how the range of entanglement changes with interactions.  We will again focus on the cases where the underlying QFT has a non-trivial three-point function.  The contributions of interest will arise from entanglement induced between a single detector in $\vec x \in A$, now with two detectors in $B$ at positions $\y_1$ and $\y_2$.  The interaction introduces a non-trivial amplitude for exciting the three detectors simultaneously, 
\beq
c_{\x,\{ \y_1, \y_2\}} =  \lambda^3 \langle \Omega | \Phi(\x) \Phi(\y_1) \Phi(\y_2) \rangle \Omega \rangle \ ,
\eeq
where we have used Equation~(\ref{eq:3_detectors}) to relate the in-in correlator and the entanglement between detectors. This contribution to the state of the detectors is illustrated in Figure~\ref{fig:entangle}. Tracing over the detectors in $B$ we arrive again at a diagonal density matrix for the single detector in $A$ with $\rho^{(A)}_{00}=1-\rho^{(A)}_{11}$, but now with
\beq
\rho^{(A)}_{11} = c_{\x,\y_1}^2 + c_{\x,\y_2}^2  + c_{\x,\{ \y_1, \y_2\}}^2 \ .
\eeq
Since the three point functions are different for classical and quantum  statistics of $\phi$, we must consider the entanglement separately as well.  First, we consider the case where $\phi$ is in the interacting (quantum) vacuum with a non-zero three-point function given by Equation~(\ref{eq:quantum_pos}).  We are specifically interested in the case where $\x$ is located near the center of the sphere and the two detectors of $B$ are close together so that $|\y_1-\y_2| \ll |\x-\y_1| \approx |\x-\y_2|$.  In this limit, we find  
\beq
c^{\rm (quantum)}_{\x,\{ \y_1, \y_2\}} \approx \lambda^3 \frac{\mu }{32\pi^4} \frac{\log |\y_1-\y_2|}{|\x-\y_1|^2} \ .
\eeq
We notice that the contribution from the interaction is suppressed relative to the contribution from the free theory.  Concretely, in  the limit $y_1, y_2 \to \infty$, we find the scaling behavior
\beq
c^{\rm (quantum)}_{\x,\{ \y_1, \y_2\}}\propto \lambda^3 \mu y_1^{-2} \ll c_{\x,\y_1} \propto \lambda^2 y_1^{-2} \ ,
\eeq
where we used $\mu \lambda \ll 1$, which is required for perturbative control.  The entanglement in the region $A$ is therefore dominated by the free theory and is again short ranged.  This is consistent with our expectations from QFT.

Now let use consider classical (excited) state for $\phi$ where the three-point correlator is given by Equation~(\ref{eq:classical_IO}).  When we compute the reduced density matrix with the detectors again near the center of the sphere, we can use Equation~(\ref{eq:classical_position}) to find
\beq
c^{\rm (classical)}_{\x,\{ \y_1, \y_2\}} \approx \lambda^3 \frac{\mu}{64 \pi^4} \frac{1}{|\x-\y_1||\y_1-\y_2|}\left(- \log \left( \frac{\theta_{\rm min}^2}{2} \right)   \, f(\hat x\cdot \hat y_{12})\right) \ ,
\eeq
where $\vec y_{12} =\y_1 -\y_2$, and $\hat x$ and $\hat y_{12}$ are the unit vectors associated with $\x$ and $\y_{12}$. We again take the limit in  the limit $y_1, y_2 \to \infty$ but instead find
\beq
c^{\rm (classical)}_{\x,\{ \y_1, \y_2\}}\propto \lambda^3 \mu y_1^{-1} |\y_1-\y_2|^{-1} \gg c_{\x,\y_1} \propto \lambda^2 y_1^{-2}
\eeq
The range of entanglement is clearly been increased by the interaction.  Furthermore, if we were to repeat the calculation in Equation~(\ref{eq:N_B}) with a large number of detectors in $B$, the integrals would no longer converge.  This long range entanglement between detectors at large distances for the classical case, which is absent in the quantum vacuum, is suggestive of the area versus volume law behavior know to distinguish the two cases.

This difference between classical and quantum correlators also manifests itself in the relative entropy~\cite{Vedral:2002zz}, a measure of distance between two states. Since we are interested in distinguishing the quantum vacuum from the classical (excited) state, consider the relative entropy for detector A, $S(\rho_{A}|\rho^{\prime}_{A}) = \text{tr}\rho_{A}\log\rho_{A}-\text{tr}\rho_{A}\log \rho^{\prime}_{A}$, where $\rho_A$ is the density matrix for detector A in the quantum vacuum and $\rho^{\prime}_A$ is the same density matrix in a classical state. Here we can write $\rho^{\prime}_A=\rho_A+\delta \rho$, where 
\beq
\delta \rho =\left(\begin{array}{cc}
(c_{\x,\{\y_1,\y_2\}}^{\rm (classical)})^2 -(c_{\x,\{\y_1,\y_2\}}^{\rm (quantum)})^2 & 0  \\
0 &-(c_{\x,\{\y_1,\y_2\}}^{\rm (classical)})^2 +(c_{\x,\{\y_1,\y_2\}}^{\rm (quantum)})^2
\end{array}\right) \ ,
\eeq
Since both $\rho_A$ and $\delta \rho$ are diagonal, $[\rho_A,\delta \rho]=0$, we can expand the relative entropy
\begin{align}
S(\rho_{A}|\rho^{\prime}_{A}) &= \text{tr}\rho_{A}\log\rho_{A}-\text{tr}\rho_{A}\log (\rho_{A}+\rho_{A}^{-1}\delta\rho-\frac{1}{2}\rho_{A}^{-2}\delta\rho^2)\\
&= \frac{1}{2}\text{tr}\rho_{A}^{-1}\delta\rho^2 \ .
\end{align}
As a result, the leading order contribution to the relative entropy is
\beq
S(\rho_{A}|\rho^{\prime}_{A})=\frac{c^{-2}_{\x,\y}}{4}((c_{\x,\{\y_1,\y_2\}}^{\rm (classical)})^2 -(c_{\x,\{\y_1,\y_2\}}^{\rm (quantum)})^2)^2 \ .
\eeq
This provides further confirmation that the difference between correlators manifests itself a robust difference in the states of the detectors used to observe them. 

\subsection{Momentum Space Detectors and Entanglement} 

Although entanglement is more often discussed in position space, momentum space entanglement~\cite{2010momentum,Balasubramanian:2011wt,Hsu:2012gk,Lello:2013bva,Peschanski:2016hgk,Grignani:2016igg} offers another useful window into the nature of cosmological correlators. Cosmological correlators are easier to calculate and represent in momentum space.  Likewise, entanglement in momentum space is easier to calculate because the vacuum of the free theory is a tensor product in the momentum basis.  As a result, entanglement between states of different momenta is uniquely a property of the interacting theory and thus is a window into the non-Gaussian nature of the correlations.  

A natural approach that combines these benefits is to work with UdW detectors that register particles in specific momentum eigenstates, rather than a specific  positions.  We can achieve this by modifying the detector response so that
\beq
\psi_i(\x) = e^{-i \k_i \cdot \x}  \ ,
\eeq
and therefore
\beq
\Phi_{i} = \lambda  \phi(\k_i,t) \ .
\eeq
By measuring momentum eigenstates at a fixed time, our detectors are responding to $\phi(\k,t)$ which is precisely what appears in our cosmological correlators.

Now we can again split our detectors into two groups, $A$ and $B$, where momenta $\vec k \in A$ and $\vec p \in B$.  We will define group $A$ as the detectors with momenta below a fixed cutoff, $k_i \leq \Lambda$, and the $B$ detectors have momenta above the cutoff, $p_j > \Lambda$.  The state of the detectors is again represented in analogy with Equation~(\ref{eq:detector_state_position}), now defining 
\begin{align}
|0,0\rangle&\equiv |0_{\{ i \}} \rangle  \\
|\{ \vec k_n \} ,0\rangle &\equiv  |1_{\k_1} ..1_{\k_n}, 0_{\{i, \hat k_{1},..,\hat k_n \}} \rangle  \\
 |0,\{ \vec p_N \} \rangle &\equiv  |1_{\p_1} ..1_{\p_N}, 0_{\{i, \hat p_{1},..,\hat p_N\}} \rangle \\
  |\{\vec k_n \},\{ \vec p_N \}\rangle &\equiv  |1_{\k_1} ..1_{\k_n},1_{\p_1} ..1_{\p_N}, 0_{\{i, \hat k_{1},..,\hat k_{n},\hat p_1,..,\hat p_{N} \}} \rangle \ .
\end{align}
As a result, the state of our momentum detectors is given by
\beq\label{eq:detector_state_momentum}
\begin{aligned}
\left|\Psi_{\rm UdW}\right\rangle=&|0,0\rangle+\sum_{\{\k_n\} \neq 0} a_{\{\k_n\}}|\{ \k_n \}, 0\rangle+\sum_{\{\p_N\} \neq 0} b_{\{\p_N\}}|0, \{p_N\} \rangle\\
&+\sum_{\{\k_n\},\{\p_N\}  \neq 0} c_{\{\k_n\}, \{\p_N\}}|\{\k_n\}, \{\p_N\}\rangle \ .
\end{aligned}
\eeq
In the free theory, the Hilbert space of $\phi$ can be decomposed into a tensor product over momentum eigenmodes via the Fock space, ${\cal H} = \otimes_{\vec p} {\cal H}_{\vec p}$.  Since each detector is tied only to a single momentum scale, the combined state of the detectors in $A$ and $B$ with the fields are similarly described by a tensor product in momentum space. More dramatically, when we project onto the vacuum of $\phi$, which is a zero momentum state, we will find the detector is never excited unless we have two detectors with equal and opposite momentum (i.e. $k$ and $-\k$ or $\p$ and $-\p$).  This reflects the fact that, in the free theory, measuring a particle at momentum $\k$ or $\p$ requires the production of an (anti)-particles with momentum $-\k$ or $-\p$. Without a second detector, this is a real particle and therefore is not in the vacuum.  Furthermore, since $k$ and $-\k$ are in $A$ and $\p$ and $-\p$ are in $B$, we cannot generate entanglement between $A$ and $B$ in the free theory.  Therefore, We need interactions in order to generate entanglement in momentum space.

We can easily determine the detector state in the interacting theory for a single detector in $A$ with momentum $\k$ and two detectors in $B$ with momenta $\p_1$ and $\p_2$.  Like the free theory, the quadratic terms vanish, $c_{k,\p_1} = c_{k, \p_2} =c_{\p_2,\p_1} = 0$, such that the leanding non-trivial contribution to the state is
\beq\label{eq:c12_momentum}
c_{\k,\{\p_1,\p_2\}} = \lambda^3 \langle  \phi(\k,t)  \phi(\p_1,t)  \phi(\p_2,t) \rangle \ .
\eeq
We can now trace over the detectors in $B$ to arrive at a diagonal reduced density matrix for $A$ with $\rho^{(A)}_{00} = 1-\rho_{11}$ and
\beq
\rho^{(A)}_{11} = |c_{\k,\{\p_1,\p_2\}} |^2 \ .
\eeq
Again, the entanglement entropy of $A$ is given in terms of this single coefficient
\beq
S^A_{\rm ent}= - {\rm Tr} \rho^{(A)} \log \rho^{(A)} = - |c_{\k,\{\p_1,\p_2\}} |^2 (\log |c_{\k,\{\p_1,\p_2\}} |^2 - 1) +{\cal O}(\lambda^8) \ .
\eeq
From Equation~(\ref{eq:c12_momentum}), we see that the entanglement entropy is determine by the in-in correlators.

Now we want to compare the nature of the momentum space entanglement for our two types of statistics. These are just our in-in correlatators in moemtnum space, so in the quantum theory we have
\beq
c^{\rm (quantum)}_{\k,\{\p_1,\p_2\}}  =-\lambda^3 \frac{\mu}{4 k p_1 p_2 (k+p_1+p_2) } \ ,
\eeq
and in the classical theory 
\beq
c^{\rm (classical)}_{\k,\{\p_1,\p_2\}}  =-\lambda^3 \frac{\mu}{16 k p_1 p_2} \left(\frac{3}{k+p_1+p_2 } +\frac{1}{k-p_1-p_2}+\frac{1}{p_1-k-p_2}+\frac{1}{p_2-k-p_1} \right) \ .
\eeq
The presence of poles at physical momenta means that $S_{\rm ent}^{(A),{\rm classical}} \gg S_{\rm ent}^{(A),{\rm quantum}}$.  This again reflects the underlying fact that creating a state with classical fluctuations introduces much stronger correlations between scales than occur in the vacuum.  

We also note that, in this case, the structure of entanglement of the UdW detectors in momentum space is proportional to the entanglement of $\phi$ in momentum space~\cite{Balasubramanian:2011wt}. The origin of this relationship is that entanglement in momentum space is trivial in the free theory.  Therefore, the perturbative construction of the detector states is consistent with the perturbative nature of the entanglement in the vacuum.  In position space, the reduced density matrix is non-trivial in the free theory and is less naturally organized as a perturbative expansion in correlators (although see e.g.~\cite{Iso:2021vrk,Iso:2021rop,Iso:2021dlj} for developments).

\section{Conclusions}\label{sec:conclusions}

The nature of the vacuum in quantum field theory is unlike any classical statistical state.  The quantum vacuum is the lowest-energy state and therefore dictates that fluctuations only have positive energies.  This fact, built into the structure of perturbative QFT, manifests itself in the structure of correlation functions and gives rise to the LSZ reduction formula relating correlators and S-matrix elements~\cite{Lehmann:1954rq}.  Classical (e.g.~thermal) fluctuations are always around a positive energy state and thus can both increase and decrease the energy.

In cosmology, these kinds of vacuum fluctuation are thought to be responsible for structure in the universe.  Yet, it remains a viable possibility that structure arose from thermal fluctuation~\cite{Berera:1995ie,Berera:1998px,Green:2009ds,LopezNacir:2011kk,LopezNacir:2012rm,Turiaci:2013dka}. One cannot perform experiments on the state of the universe to isolate their quantum nature and resolve this question~\cite{Maldacena:2015bha}.  Instead, we must rely on statistical properties of the initial conditions to infer how structure was created.  Concretely, it was proposed in~\cite{Green:2020whw} that the difference in the analytic structure of correlators for quantum and classical fluctuations is both completely general and observable.  

In this paper, we demonstrated that this proposed cosmological Bell-like test has a flat space analogue.  The same analytic structure seen in cosmological correlators appears in both in-in and in-out correlators in flat space, and is responsible for the LSZ reduction for quantum vacuum correlators.  For classical correlators, the additional poles seen in inflationary correlators is a direct consequence of scattering processes involving particles that are necessarily present in the initial state. 

The meaning of the in-in correlator is less clear in flat space than it is in cosmology.  In cosmology, we interpret these correlators as a signal of cosmological particle production.  In flat space, the interacting (quantum) vacuum is a well-defined energy eigenstate  containing no particles and, yet, has non-zero in-in correlations.  Instead, we show the flat space in-in correlator contributes to the amplitude for exciting a localized (Unruh - de Witt) particle detector.  The particle production arises in flat space because of the uncertainty principle: a particle detector that is sufficiently localized in space and time to make such a measurement breaks translations and excites particles from the vacuum.  We additionally show that one can use the entanglement of these detectors as probes to the entanglement of the underlying field.

Much is known about the unique properties of quantum mechanical systems in flat space.  Naturally one would hope these insights could be applied to cosmology, particularly in light of some analogous structure of the correlators.  The central challenge is that cosmological observables are classical, for all practical purposes; one cannot simply expose the quantum nature of cosmology through the direct measurement of non-commuting observables.  Yet, the question of what kinds of initial conditions can only be prepared in a quantum universe is closely related to the problem of quantum state preparation on a quantum computer. One might hope that intuition from quantum computing could shed further light on cosmology, or vice versa~\cite{Swingle:2014qpa,Swingle:2016foj}.

\paragraph{Acknowledgements}

We are grateful to Daniel Baumann, Jonathan Braden, Dick Bond, Tim Cohen, John McGreevy, Mehrdad Mirbabayi, Eva Silverstein, and Rafael Porto for helpful discussions. The authors were supported by the US~Department of Energy under grant no.~\mbox{DE-SC0009919}.

\appendix

\section{Classical Hidden Variables and Anti-particles}\label{app:nonlocal}

In our classical in-in correlators, the appearance of poles at physical momenta can be traced to the form of the causal propagator, 
\beq
G(k,t) = \frac{\sin (k t)}{k} = \frac{1}{2 i k}
\left(e^{i k t} - e^{-i k t} \right) .
\eeq
The appearance of both positive and negative frequency modes in this expression is responsible for the additional poles in the classical correlators.  If we were to simply remove the negative frequency term in this expression in the limit $t\to 0$ the propagator is non-analytic in $k$, $G(k, t=0) \propto 1/k$. This would imply a violation of causality as the propagator is non-zero at space-like separated points. This is the precise sense in which replicating the quantum signal requires a non-local theory~\cite{Bell:1964kc,Bohm:1951xw,Bohm:1951xx}.

The need for a negative frequency mode in the causal propagator is essentially the same reason that causality demands anti-particles. To see this connection, we will work with a complex scalar, $\varphi$, such that 
\beq
\varphi(\x,t) = \int \frac{d^3 k}{(2\pi)^3} \frac{1}{\sqrt{2 k}} e^{-i \k\cdot \x} [ a_\k  e^{-i k t} + b^{\dagger}_{-\k}  e^{i k t}  ] \ .
\eeq
In contrast to a real scalar, we have the option of setting $b_{-\k} = 0$ so that we propagate only positive frequency modes.  

To make sense of the classical case, we will consider the quantum theory at large occupation number, as we did in Section~\ref{sec:inin}.  Suppose the field $\phi$ is in an $n$-particle and $n$-anti-state at each momenta, as defined in Equation~(\ref{eq:n_state}).  Schematically, the field would act on this state as
\begin{align}
\varphi(\k,t) |n,n\rangle  &\to A e^{-i k t} |n_\k-1,n\rangle + B^* e^{i k t}|n,n_{-\k}+1\rangle \\
\bar\varphi(\k,t) |n,n\rangle  &\to A^* e^{i k t} |n_{-\k}+1,n\rangle + B e^{-i k t} |n,n_{\k}-1\rangle \\
\langle n , n| \varphi(\k,t)   &\to  \langle n,n_{\k}-1 | B^* e^{i k t} +  \langle n_{-\k}+1,n | A e^{-i k t} \\
 \langle n, n | \bar\varphi(\k,t)  & \to  \langle n,n_{-\k}+1 | B e^{-i k t} +  \langle n_{\k}-1,n |A^* e^{i k t}
\end{align}
where $|n,m \rangle$ is an $n$ particle $m$ anti-particle state, and $A$ and $B$ are time-independent functions of $k$ that are determined by how $a_\k$ and $b_\k$ act on these states. Concretely, if $B=0$ it would imply $b$ and $b^\dagger$ annihilate the state $|n,n\rangle$.

Using how the operators act on the states, we can write the two point functions as 
\beq
\langle  \varphi(\k,t) \bar \varphi(\k',t') \rangle =  |B|^2 e^{ik(t-t')}   \langle n,n_{\k}-1 |n, n_{-\k'}-1\rangle +  |A|^2 e^{-ik(t-t')}   \langle n_{\k'}+1,n |n_{-\k}+1, n\rangle
\eeq
\beq
\langle \bar \varphi(\k,t)  \varphi(\k',t') \rangle  =  |A|^2 e^{ik(t-t')}  \langle n_{\k}-1,n|n_{-\k'}-1 , n\rangle + |B|^2 e^{-ik(t-t')} \langle n,  n_{-\k}+1 |n, n_{\k'}+1\rangle  .
\eeq
In the local model $A=B=1/(\sqrt{2 k} )$.  We argued from the structure of the causal Green's function that $B=0$ is nonlocal, which eliminates the possibility of locally propagating only the positive frequency modes classically. 

We can see the connection between $B=A$, locality, and antiparticles using time-order (in-out) correlators. We start by noting that $a^{\dagger}$ creates a positive energy state when acting on the right and negative energy state when acting on the right.  However, since the time ordering can be changed by a Lorentz boost, $\varphi$ must create a positive energy state  when acting on either the right or the left.  As a result,  $\varphi$ creates a particle when acting on the right and an anti-particle when acting on the left.  This is only possible with $|A|=|B|$.  

\phantomsection
\addcontentsline{toc}{section}{References}
\bibliographystyle{utphys}
\bibliography{QRefs}

\end{document}